# Deformation and dynamics of erythrocytes govern their traversal through microfluidic devices with a deterministic lateral displacement architecture

Wei Chien[a], Zunmin Zhang[a], Gerhard Gomppe[a], and Dmitry A. Fedosov[a]

*[a] Theoretical Soft Matter and Biophysics, Institute of Complex Systems and Institute for*

*Advanced Simulation, Forschungszentrum Jülich, 52425 Jülich, Germany*

Deterministic lateral displacement (DLD) microfluidic devices promise versatile and precise processing of biological samples. However, this prospect has been realized so far only for rigid spherical particles and remains limited for biological cells due to the complexity of cell dynamics and deformation in microfluidic flow. We employ mesoscopic hydrodynamics simulations of red blood cells (RBC) in DLD devices with circular posts to better understand the interplay between cell behavior in complex microfluidic flow and sorting capabilities of such devices. We construct a mode diagram of RBC behavior (e.g. displacement, zig-zagging, and intermediate modes) and identify several regimes of RBC dynamics (e.g. tumbling, tank-treading, and trilobe motion). Furthermore, we link the complex interaction dynamics of RBCs with the post to their effective cell size and discuss relevant physical mechanisms governing the dynamic cell states. In conclusion, sorting of RBCs in DLD devices based on their shear elasticity is in general possible, but requires fine-tuning of flow conditions to targeted mechanical properties of the RBCs.

## I. Introduction

Sorting biological cells based on their intrinsic properties and the detection of a cell subpopulation with specific mechanical characteristics are important objectives directly relevant for a number of biomedical and clinical applications.[1-3] The capability of sorting erythrocytes or red blood cells (RBCs) based on changes in their elastic properties is relevant for several blood diseases and disorders, such as sickle-cell anemia and malaria.[4,5] Deterministic lateral displacement (DLD) microfluidic devices have demonstrated excellent and precise performance for size-based label-free separation of rigid spherical colloidal particles.[6-8] DLD devices employ a designed array of obstacles (or posts) sandwiched between two planar walls, arranged such that every row of posts is laterally shifted with respect to the previous row. Such geometric arrangement leads to several laminar flow streams within the DLD device, where the fluid stream adjacent to a post (also called the first stream) passes under the following post,[9,10] see Fig.1. The thickness $\beta$ of the adjacent stream defines the critical size of the device, such that particles with a smaller size than $2\beta$ traverse the DLD device in a zig-zagging (ZZ) fashion, following the streamline with a nearly zero lateral displacement, while larger particles are bumped laterally by consecutive shifted posts, switching the flow lanes in a displacement (DP) mode. Recent studies [8, 11-13] have also demonstrated the existence of a mixed mode for rigid spherical particles with a lateral displacement between the ZZ and DP modes. Thus, in this ratchet-like way, precise sorting of rigid spherical particles based on their size can be achieved.

DLD devices have also been applied to sorting biological cells.[14-19] The separation results are generally good when biological particles of interest possess significantly different sizes, such as erythrocytes and leukocytes in the blood, but are poor when targeted cell populations have overlapping sizes. In comparison to rigid spherical particles, biological cells are deformable and often have a non-spherical shape. For example, RBCs have a biconcave disk-like shape and are subject to strong and dynamic deformations in fluid flow.[20-23] Therefore, biological cells deform in response to local fluid stresses in DLD devices, and thus cannot be straightforwardly characterized by a fixed size.[16, 24, 25] Instead, the lateral displacement of biological cells in DLD devices can be characterized by an effective size, which is different from their undeformed size and depends on the cell's mechanical properties and local fluid-flow stresses within the device.

In the context of sorting RBCs, several studies[17, 24, 25] have considered "thin" DLD devices, for which the height $h$ is smaller than the RBC diameter of $8\ \mu m$. In such devices, the rotational freedom of RBCs is constrained between the lower and upper walls, simplifying the cell dynamics within the device. Even though thin DLD devices have shown some sensitivity to RBC deformability,[17, 24] they are prone to clogging due to frequent interactions of RBCs with the walls. In contrast, "thick" DLD devices with $h > 8\ \mu m$ allow for a much higher throughput, but do not put any restrictions on the RBC's orientation, yielding a more complex cell dynamics. An experimental study of RBC traversal through a thick DLD device has shown a poor sensitivity to RBC deformability,[25] because a RBC orients itself perpendicular to the shear plane, exposing its thickness ($\approx 2.88\ \mu m$) for sorting instead of its long diameter ($\approx 8\ \mu m$). However, a recent experimental and simulation investigation[16] has demonstrated RBC sorting in a thick DLD device based on the viscosity contrast $C = \eta_{in}/\eta_{out}$ between RBC cytosol and suspending medium, emphasizing the importance of RBC dynamics in the device. Flipping dynamics of RBCs has also been hypothesized to play an important role in their traversal through DLD devices with I-shaped posts.[19] Furthermore, a recent numerical study suggests that sharp-edged DLD posts are advantageous for deformability-based RBC sorting, as they impose pronounced cell-deformation modes.[26] The importance of both RBC dynamics and deformation in DLD flow for deformability-based sorting demands a better understanding of the interplay between these two characteristics and their effect on sorting of RBCs based on their intrinsic properties.

To this end, we employ a thick DLD device with circular-shaped posts and investigate sorting of RBCs based on their elastic properties at a viscosity contrast $C = 1$. We construct a mode diagram of RBC behavior for a wide range of membrane shear elasticities $\mu$ and row shifts $\Delta\lambda$, which allows us to characterize the dependence of the DP-ZZ transition on $\mu$ or the corresponding capillary number $Ca$. In particular, we identify three different regimes of RBC behavior depending on $Ca$. At low $Ca$, RBCs exhibit tumbling (TU) dynamics with small shape deformations, while at large $Ca$, RBCs generally show tank-treading (TT) dynamics with significant stretching in the flow direction. In the intermediate $Ca$ regime, RBC dynamics is characterized by a superposition of TU and TT motion with strong and dynamic shape deformations. These dynamic states are qualitatively similar to those of RBCs in simple shear flow.[21-23, 27, 28] Interestingly, the DP-ZZ transition is a non-monotonic and re-entrant function of $Ca$, which can be explained by an effective RBC size within the DLD device. As a result, we provide a link between the effective cell size and dynamic RBC behavior, and identify the physical mechanisms involved.

The paper is organized as follows. The employed models and methods including simulation setup are described in Section II. Section III presents the mode diagram for a wide range of capillary numbers and a detailed analysis of dynamic RBC behavior in the DLD device. Finally, in Section IV, we summarize and discuss the simulation results and the importance for RBC sorting.

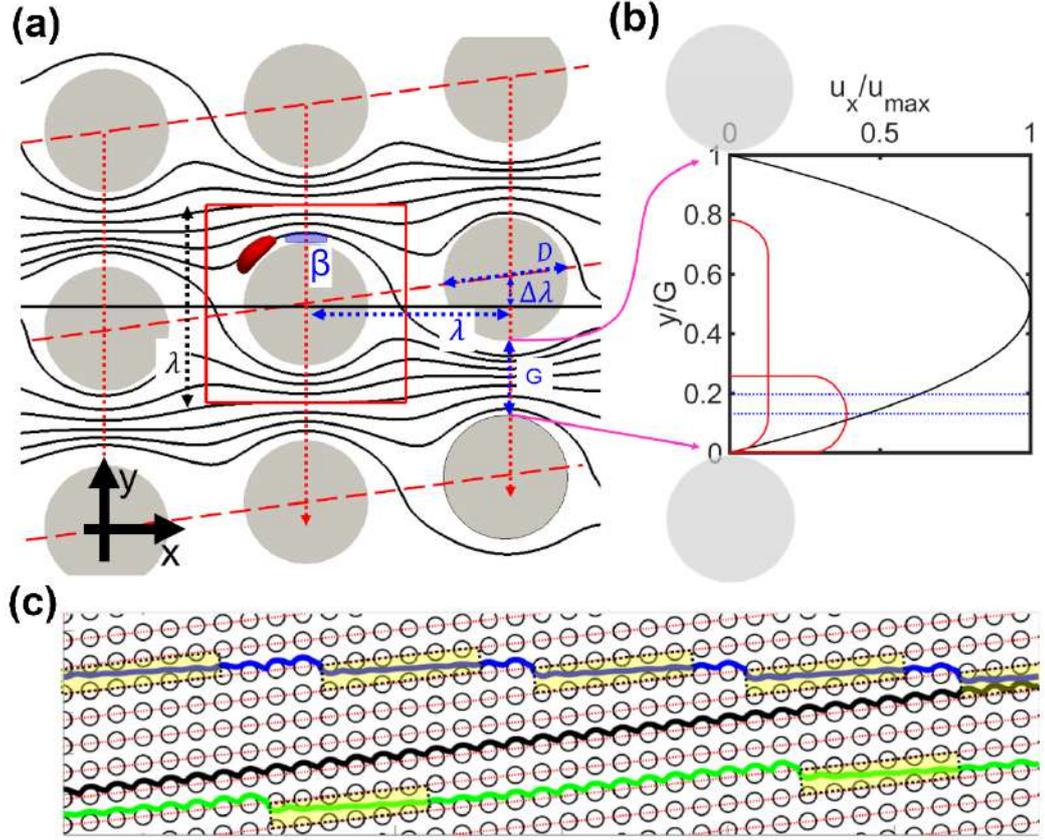

FIG. 1. Schematic of a DLD device and cell traversal through it. (a) Geometrical parameters of the device, including the post distance $\lambda = 35$ μm, the gap size $G = 10$ μm, and the row shift $\Delta\lambda$. The red square indicates the simulation domain. $\beta$ shows the critical size of the device determined from the thickness of the first stream. (b) Schematic velocity profile within the gap between two posts. The physical contours of a RBC are shown in red. The blue dashed lines indicate two critical sizes of the first flow stream: $\beta_{min}(\Delta\lambda = 2.4\text{μm}, \Delta\lambda/\lambda = 0.07)$ and $\beta_{max}(\Delta\lambda = 3.6\text{μm}, \Delta\lambda/\lambda = 0.1)$, discussed in Section III.C. (c) Illustration of several trajectories of a RBC within the DLD device with n = 8, representing displacement (DP) mode (black line), zigzagging (ZZ) mode (blue line), and intermediate (IM) mode (green line). The boundaries of the device lanes are shown by the red dashed lines. The yellow stripes mark the portions of the RBC trajectories, where the RBC does not directly interact with the posts, so that these portions are irrelevant for the selection of DP or ZZ motion.

## II. Models and methods

In simulations, a coarse-grained RBC model[29-32] is employed and coupled to a mesoscopic representation of fluid flow by the smoothed dissipative particle dynamics method.[33-36] The simulation approach is similar to previous work,[16] where a good quantitative agreement between simulations and corresponding experiments has been achieved.

### A. RBC membrane model

The cell membrane is modeled by a triangulated surface[29-31] with 1000 vertices, 1996 faces, and 2994 bonds to have a resolution around 0.4 μm, which is the average bond length. We use a RBC model with average size and shape (see Table I) characterized by an effective diameter $D_r = \sqrt{A/\pi}$ and reduced volume $v_r = 6V/(\pi D_r^3)$, where $V$ is the cell volume and $A$ the surface area. Potential energies are applied to capture the mechanical properties of the membrane including the in-plane elastic energy which mimicks elasticity of the spectrin network, the bending energy which represents bending rigidity of the lipid bilayer, and the constraints for area and volume which account for the incompressibility of the lipid bilayer and inner cytosol, respectively.[29, 30] This model has been shown to capture RBC deformation even in the nonlinear regime.[29, 30, 37] The mechanical properties quantified by the shear modulus µ and the bending rigidity $\kappa$ are also summarized in Table I. The dimensionless Föppl-von-Kármán number $\Gamma = \mu D_r^2/\kappa$ is chosen to mimic that of a healthy RBC.[38] The RBC model assumes a stress-free shape of the elastic network to be spheroidal with $v_r^{free} = 0.96$. Furthermore, we apply a homogeneous spontaneous curvature $\tilde{c}_0$ characterized by the normalized curvature $c_0 = \tilde{c}_0\, D_r/2 = 3$, following a previous numerical study.[39] The strengths of the total area, local area, and volume constraint coefficients ($k_a$, $k_d$, and $k_v$) are set large enough to restrict the total change in the surface area and volume to be within 5% for all simulated flow conditions.

TABLE I: The standard parameters of a healthy RBC used in this study: the cell area $A$, cell volume $V$, cell diameter $D_r$, reduced volume $v_r$, bending rigidity $\kappa$, the shear modulus $\mu$, and Föppl-von-Kármán number $\Gamma$.

| | | | |
|---|---|---|---|
| $A$ | 132.90 µm² | $\kappa$ | 70$k_B$T |
| $V$ | 90.21 µm³ | $\mu$ | 4.67 µN/m |
| $D_r$ | 6.51 µm | $\Gamma$ | 681.5 |
| $v_r$ | 0.62 | | |

## B. Modeling fluid flow in a DLD device

The fluids inside and outside the cell are modeled by the smoothed dissipative particle dynamics method (SDPD) with angular momentum conservation.[34] SDPD is a particle-based hydrodynamics method that is obtained by a Lagrangian discretization of the Navier-Stokes equations[33-36] and includes consistent thermal fluctuations.[35] In SDPD, each particle represents a small fluid volume and interacts with other surrounding particles via pairwise soft potentials. The dynamic viscosity of a SDPD fluid is set directly as an input parameter. SDPD parameters are selected such that the simulated fluid can be considered incompressible[40] for Mach numbers $Ma = u_{max}/u_s < 0.01$, where $u_{max}$ is the maximum flow velocity and $u_s$ is the speed of sound.

The coupling between the SDPD fluid and the membrane particles is implemented by viscous friction through a dissipative force.[30] Bounce-back reflection of fluid particles is applied at the membrane triangles to impose membrane impenetrability. The viscosity of the internal fluid, $\eta_{in}$, is set to be the same as that of external fluid, $\eta_{out} \equiv \eta_{in}$, so that the viscosity ratio $C = \eta_{in}/\eta_{out}$=1. Solid walls are modeled by a layer of immobile SDPD particles, whose structure and density are identical to those of the SDPD fluid. Therefore, interactions between fluid and solid particles are the same as fluid-fluid interactions. The particle number density is set to $\rho = 9\ \mu m^{-3}$ to reasonably resolve fluid flow within 0.48 $\mu m$, which is comparable to the resolution of the membrane. In addition, a reflective boundary condition is applied to fluid and membrane particles at solid surfaces and an adaptive shear force is added to fluid particles in the direction tangential to a solid surface in order to ensure no-slip boundary conditions.[41]

## C. Simulation setup

Figure 1a shows a typical DLD geometry with cylindrical posts of circular cross-section. The employed geometry is characterized by post distance $\lambda$ = 35 μm, post diameter D = 25 μm, gap size G = 10 μm, and row shift $\Delta\lambda$ that is varied. The corresponding geometric period of the DLD device is $n = \lambda/\Delta\lambda$. The height of the DLD device or the distance between two confining surfaces is set to $h = 10$ μm, such that a RBC has full rotational freedom within the device. The simulation domain contains a square region with a single cylindrical post that is indicated by the red square in Fig. 1a. In both flow (x) and gradient (y) directions perpendicular to the post axis, periodic boundary conditions are employed. However, for the periodic boundary in the x direction a row shift $\Delta\lambda$ is introduced along the y direction to represent the shift of subsequent rows of posts. The flow is driven in the x direction by a constant body force applied to each fluid particle. In addition, an adaptive force is applied in the y direction to ensure no net flux in the flow gradient direction.[8, 16]

To characterize the flow stresses in comparison to cell deformation, the capillary number Ca = $\dot{\gamma}\eta D_r/\mu$ is employed with an average shear rate $\dot{\gamma} = u_{max}/G$, where $u_{max}$ is the maximal value in the parabolic velocity profile captured from simulations without RBCs, as shown in Fig. 1b. Ca is altered by changing the membrane shear modulus μ. Other mechanical parameters of a RBC (e.g. $\kappa$, $k_a$, $k_d$, and $k_v$) are also re-scaled together with μ by the same factor, so that the Föppl-von-Kármán number Γ remains fixed and the constraints of cell area and volume are consistent. Furthermore, the

Mach number Ma and Reynolds number $\text{Re} = \rho G u_{max}/\eta < 0.3$ of the flow are unaffected by this variation of Ca. The capillary number of course also depends linearly on the flow rate (or applied pressure gradient) within the DLD device. We employ the definition of Ca based on the shear modulus, since µ is likely a dominant factor for RBC dynamics.[21] However, another capillary number $Ca_\kappa = \dot\gamma \eta D_r^3/\kappa$ based on the membrane bending rigidity may also affect RBC dynamics. Note that Ca and $Ca_\kappa$ are related through the Föppl-von-Kármán number $\Gamma$ that is kept constant in our investigation.

## III. Results

To characterize the lateral displacement of a RBC along the y-direction, we compute an average lateral shift $l_{ave}$ per row of posts by monitoring RBC trajectories. Then, the displacement index $I_d = l_{ave}/\Delta\lambda$ is a convenient measure for distinguishing various types of RBC lateral displacements. In simulations, the motion of a RBC is classified as the DP mode whenever $I_d > 0.8$ (black trajectory in Fig. 1c), such that the cell nearly always remains within the same device lane (or between two red dashed lines in Fig. 1c). The ZZ mode has $I_d < 0.2$, where the RBC switches device lanes for every n or n+1 rows and attains nearly zero net lateral displacement (blue trajectory in Fig. 1c). Trajectories with $0.2 \leq I_d \leq 0.8$ are classified as the intermediate (IM) mode, where the cell requires a much longer distance (beyond n+1 rows) to cross a single device lane, as indicated by the green trajectory in Fig. 1c.

### A. DLD mode diagram

By measuring $I_d$ for various row shifts $\Delta\lambda$ and capillary numbers Ca, we can construct a DLD mode diagram for a single RBC, as shown in Fig. 2. We focus mainly on the transition between DP and ZZ modes, which can be characterized by a small range of $\Delta\lambda$ values for every fixed Ca, because it represents a good separatrix for sorting. Interestingly, this transition or a corresponding critical shift $\Delta\lambda_c$ (marked schematically by the red dashed line in Fig. 2) is a complex function of Ca. For Ca $\leq$ 0.01, $\Delta\lambda_c$ varies within the range 3.2 - 3.6 µm, and reaches a maximum at Ca = 0.01. For 0.01 < Ca < 0.3, the critical shift first decreases to a minimum value around Ca $\cong$ 0.05, and then increases again for larger Ca values. Finally, for Ca $\geq$ 0.3, $\Delta\lambda_c$ continues to slightly increase but seems to level off at large Ca values. In fact, the regions of Ca, in which $\Delta\lambda_c$ has the strongest variations, are the most interesting for deformability-based sorting, because cells with differing shear moduli would attain distinct traversal modes through the DLD device. Therefore, flow conditions within a DLD device have to be selected carefully and should be directly associated with the targeted mechanical properties of RBCs. Note that the range of capillary numbers in Fig. 2 covers about three orders of magnitude, representing a large range of RBC shear moduli or similarly a large range of flow rates within the DLD device.

Figure 3 shows the average RBC velocity $u_{RBC}$ normalized by the average fluid velocity $U_{ave}$ as a function of $\Delta\lambda$ and Ca. Interestingly, the transition between DP and ZZ modes is strongly correlated with $u_{RBC}/U_{ave}$, such that this ratio is generally smaller than unity near the transition. This means that a RBC in the IM mode moves on average slower than the average flow velocity, while the DP and ZZ modes generally lead to $u_{RBC}/U_{ave} > 1$. In the IM mode, RBC often comes close to the stagnation point at the left-hand side of posts, where flow nearly vanishes. A similar behavior of RBCs has been also found in a thin DLD device, whose height is smaller than the RBC diameter.[24] Note that the width of the transition region with $u_{RBC}/U_{ave} < 1$ is larger for low Ca than for high Ca values.

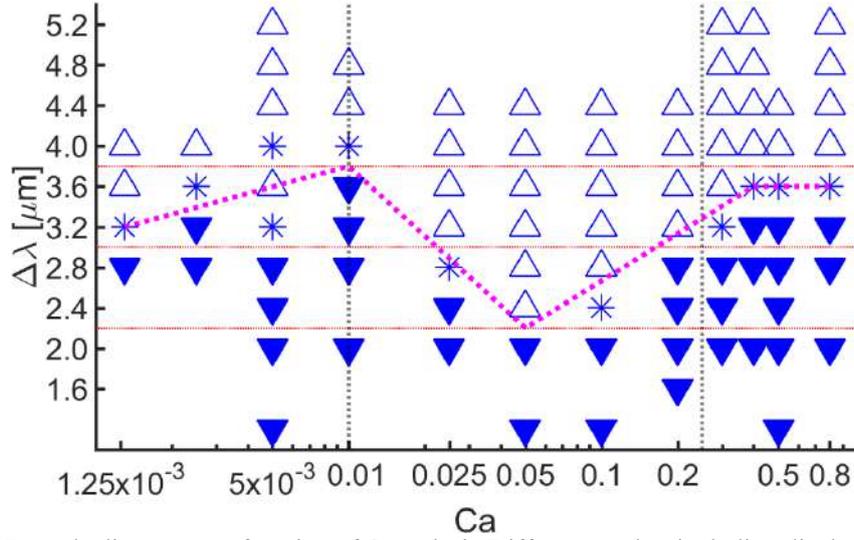

FIG. 2. DLD mode diagram as a function of Ca and Δλ. Different modes, including displacement (DP, solid triangles), intermediate (IM, * symbols), and zigzagging (ZZ, hollow triangles) modes, are shown. The resolution in Δλ is 400nm. The DP-ZZ transition or a corresponding critical shift $\Delta\lambda_c$ is indicated schematically by the red dashed line. $I_d$ at each point is calculated for RBC traversal over at least $2n$ rows.

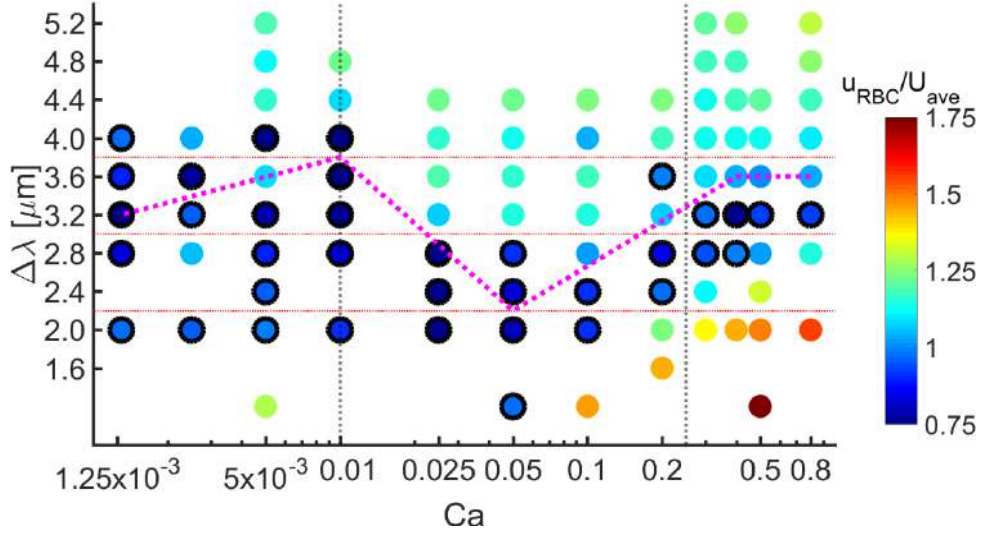

FIG. 3. Color map of RBC velocity $u_{RBC}$ normalized by the average flow velocity $U_{ave}$ as a function of Ca and $\Delta\lambda$. The black circles mark the conditions for which the ratio $u_{RBC}/U_{ave}$ is smaller than unity. The DP-ZZ transition or a corresponding critical shift $\Delta\lambda_c$ is indicated schematically by the red dashed line.

## B. Dynamic states of RBCs in DLD devices

Clearly, the dependence of $\Delta\lambda_c$ on capillary number Ca, marked schematically by the dashed line in Fig.2, is due to RBC dynamics and deformation generated by the fluid stresses near the posts. To better understand the dependence of the DP-ZZ transition on Ca, we characterize the deformation and dynamics of RBCs by several quantities, see Fig. 4. These include the inclination angle θ and the tilt angle Φ (Fig. 4a), where θ is the angle between the cell's orientational vector $\bm{n}$ and the x axis within the x-y shear plane (i.e. projected onto the plane), and Φ is the angle between $\bm{n}$ and the x-y plane, characterizing the tilt of the RBC from the shear plane. Note that the orientational vector $\bm{n}$ is determined by the eigenvector of the inertia tensor corresponding to the smallest eigenvalue. Furthermore, we monitor the time-dependent motion of the membrane through a tank-treading angle $\theta_D$ measured in a reference frame attached to the eigenvectors of the inertia tensor for a material region initially located at the RBC dimple (Fig. 4b). At the beginning of each simulation $\theta_D|_{t=0} = 0$, and deviates from the initial value whenever the RBC membrane is set into a tank-treading-like motion by fluid stresses. RBC deformation is characterized by the largest cell length $a$ and the relative lateral stretching $W_{eff} = bc/(b_0 c_0)$, where $a \geq b \geq c$ are calculated from the eigenvalues $\lambda_i$ of the inertia tensor (Fig. 4c), with $a = \sqrt{5(\lambda_2 + \lambda_3 - \lambda_1)/2}$ , $b = \sqrt{5(\lambda_3 + \lambda_1 - \lambda_2)/2}$ , and $c = \sqrt{5(\lambda_1 + \lambda_2 - \lambda_3)/2}$. The biconcave rest shape of a RBC has $a_0 = 4.277 > b_0 = 4.272 > c_0 =$

1.442 [μm]. Even though $W_{eff}$ is mainly intended to describe the lateral width of the RBC near the posts, it can also be used to distinguish different types of RBC deformation within DLD devices.

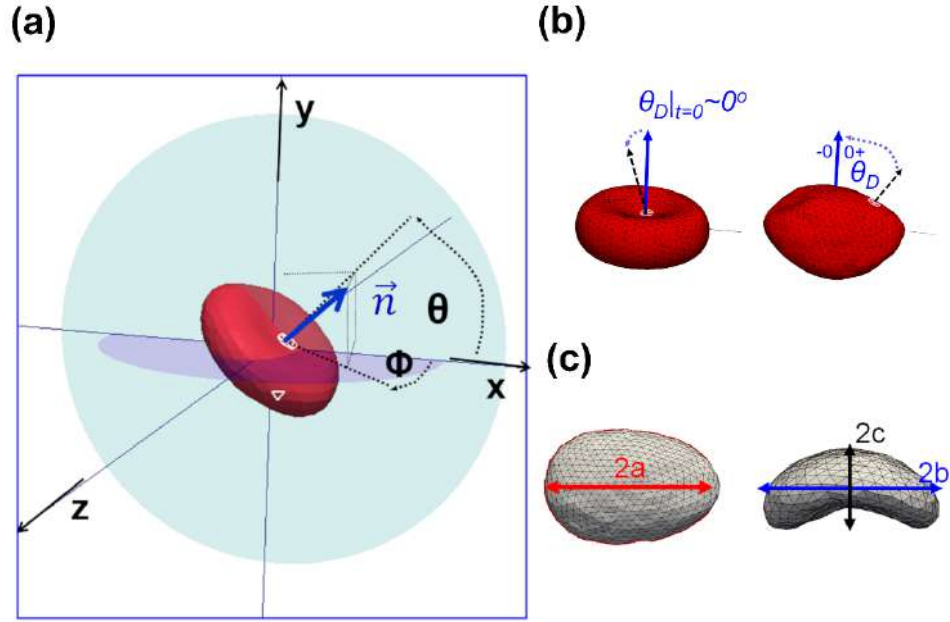

FIG. 4. Different quantities to characterize dynamical states of a RBC. (a) Cell orientation described by the inclination angle θ with respect to the x-axis within the x-y plane (flow-gradient plane) and the tilt angle Φ from the x-z plane. The cell's orientational vector $\boldsymbol{n}$ is the symmetric axis of the shape drawn in blue. (b) Membrane motion characterized by a tank-treading angle $\theta_D$. The white patch denotes a moving material region whose initial position corresponds to the cell's dimple and the orientaional vector $\boldsymbol{n}$ is marked by the blue arrow. (c) Deformation of a RBC characterized by its largest length $a$ and the relative lateral stretching $W_{eff} \equiv bc/(b_0 c_0)$, where $b$ and $c$ represent the lateral sizes of the RBC.

## RBCs in a displacement mode

Large portions of ZZ and IM trajectories are not very important for the determination of lane crossing, as marked with yellow stripes in Fig. 1c. They are identical and have essentially the same length. The RBC just flows along the stream within a single device lane without direct interaction with the posts. The cell approaches the post surface eventually, and the interaction between the RBC and the post surface is what matters, but it occurs less frequently in ZZ and IM modes than in the DP mode. Thus, we focus on the dynamics in DP, where the cell is locked in a certain device lane and repeatedly collides with the circular posts.

Along a DP trajectory, RBCs experience periodic and non-uniform local shear stresses, which are generally largest near the posts. Therefore, RBCs in the DP mode are subject to a periodic stretching and relaxation process in DLD flow, depending on the flow strength and RBC mechanical properties. In addition to RBC deformation, cell dynamics may affect the effective size. The flow near a post can be approximated as a linear shear flow, in which a RBC is known to exhibit tumbling (TU) for low Ca and tank-treading (TT) for high Ca, at low viscosity contrasts $C \leq 3$. [21-23, 27, 42, 43] The main mechanism of this transition is the shape memory of the cell's elastic network,[28] and is related to an energy barrier for the TT motion of a RBC membrane.[42, 43] Thus, fluid stresses are too small to overcome the energy barrier at low Ca, resulting in a TU motion of the RBC in simple shear flow. At high Ca, fluid stresses can overcome the elastic energy barrier, and RBC membrane sets into TT motion. For intermediate Ca between the TU and TT modes, there also exists a rolling motion[21, 27] characterized by RBC orientation perpendicular to the shear plane, i.e. $\Phi \cong 90°$. In our DLD simulations, the rolling motion is nearly absent. The tilt angle relaxes to zero eventually for a wide range of Ca values, see Fig. 5. This is likely due to multiple collisions of RBCs with the posts in DP mode, which favors a cell's orientation with $\Phi \approx 0°$ for subsequent posts. Thus, we assume the initial cell orientation of $\Phi = 0°$ in all simulations.

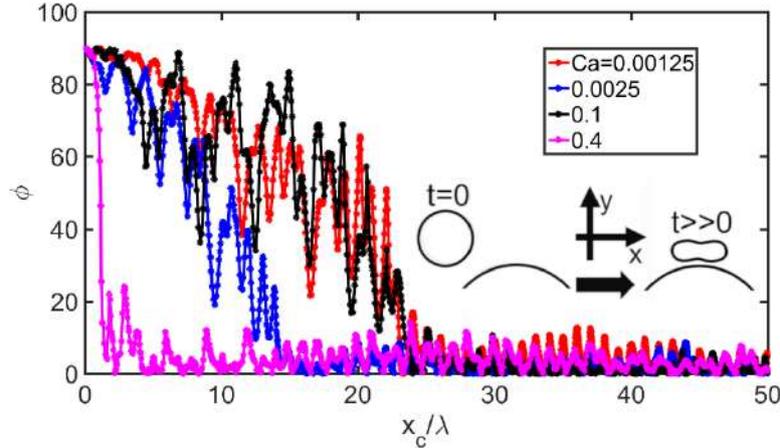

FIG. 5. Relaxation of the tilt angle $\Phi$ with an initial RBC orientation of $\Phi|_{t=0} = 90°$. In all simulations, the RBC attains a DP mode in a DLD device with $\Delta\lambda = 2$ μm, $\Delta\lambda/\lambda = 0.057$. The orientation change is schematically shown in the inset.

## Dynamic cell states at low capillary numbers

At low Ca ∈ [0.001; 0.01], flow stresses are too weak to overcome the membrane's elastic barrier for tank-treading and the RBC tumbles similarly to a solid disc. In particular, the RBC slides along the curved post surfaces and tumbles after passing the top of the posts, as illustrated in Fig. 6a and confirmed quantitatively by the time evolution of the inclination angle θ that spans 360 degrees (Fig. 6c). The tilt angle remains close to $\Phi \approx 0°$ over the whole trajectory, independently of Ca. The tank-treading angle $\theta_D$ in Fig. 6b shows that the RBC membrane is subject to moderate oscillations whose amplitude becomes larger with increasing Ca. These oscillations indicate that the flow stresses initiate TT motion, but are too small to impose full tank-treading. Figures 6d-e show that the RBC experiences small deformations with relative lateral stretching $W_{eff}$ increasing by about 10% (Fig. 6e). The increase in $W_{eff}$ corresponds to the shape change from a biconcave shape with two dimples to a bowl shape with a single dimple, as illustrated in the inset of Fig. 6a, and in Movie S1 for Ca = 0.01. In conclusion, RBCs at low Ca exhibit TU dynamics with small membrane deformations in the DLD device, nearly independent of capillary number.

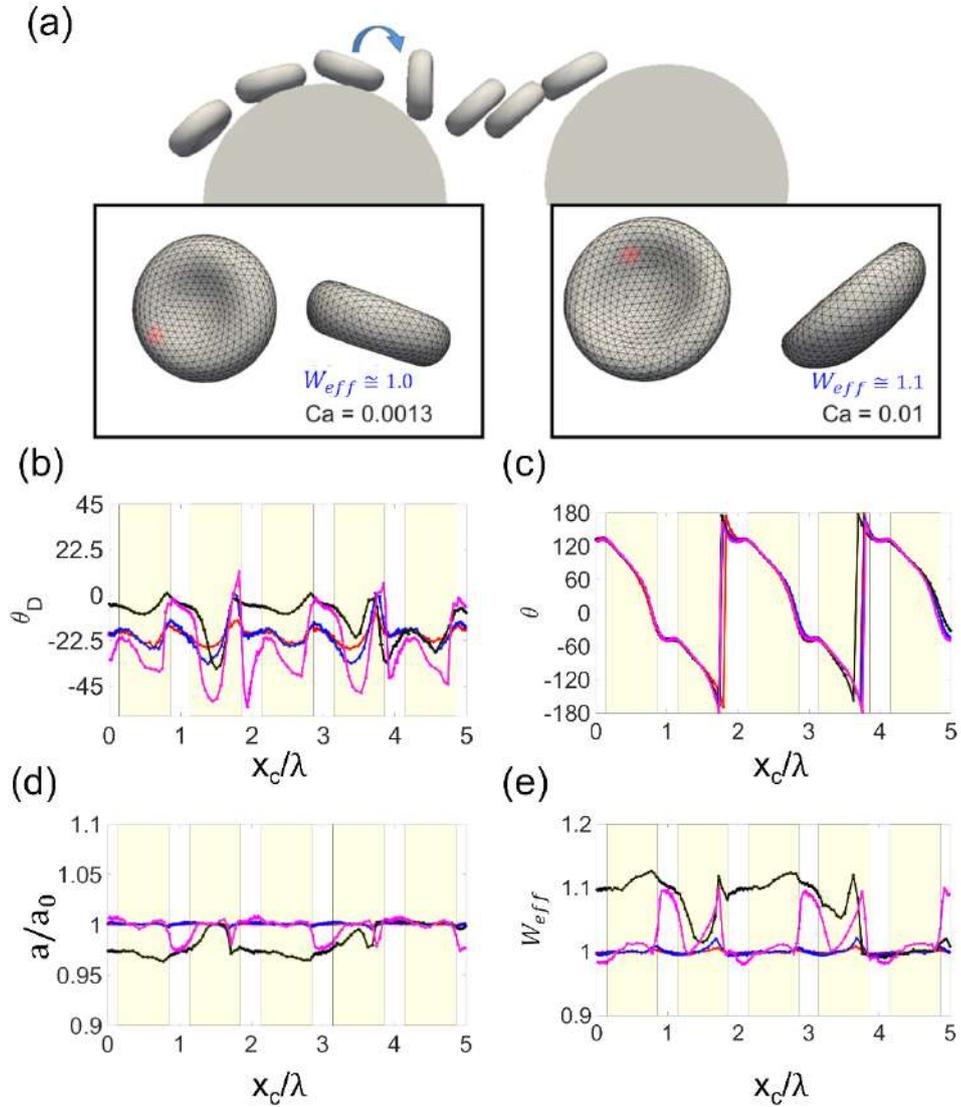

**FIG. 6.** Characteristics of tumbling-dominated displacement modes of RBCs in the DLD device with $\Delta\lambda = 2.0$ μm, $\Delta\lambda/\lambda = 0.057$. (a) Snapshots of tumbling-dominated motion at Ca = 0.0013 (see also Movie S1 for Ca = 0.01). The cell shapes with $W_{eff} \approx 1$ and $W_{eff} \approx 1.1$ are shown below. Evolution of (b) tank-treading angle $\theta_D$, (c) inclination angle $\theta$, (d) RBC stretching characterized by $a/a_0$, and (e) relative lateral stretching $W_{eff} = bc/(b_0 c_0)$. Colors represent capillary numbers: Ca = 0.00125 (red), 0.0025 (blue), 0.005 (black), 0.01 (pink). Yellow shaded regions indicate the post locations.

## Dynamic cell states at intermediate capillary numbers

At intermediate Ca ∈ [0.01; 0.15], a RBC in the DLD device with $\Delta\lambda$ = 2.0 µm does not perform pure TU or pure TT motion, since neither θ nor $\theta_D$ span 360 degrees, as shown in Figs. 7c-d. The RBC motion is accompanied by strong and dynamic membrane deformations sensitive to $Ca$. A frequently occurring deformed shape is the trilobe (Tri) [21, 44] (Figs. 7a-b, Movie S2 and Movie S3) with a relative lateral stretching larger than $W_{eff} \approx 1.2$. Similarly as in the low-Ca regime, $W_{eff} \approx 1.1$ represents a bowl shape, which also occurs when the RBC is located between two posts. Strong membrane deformations, especially for the trilobe shape, imply abrupt jumps of the cell's major axes, leading to discontinuous behaviour of the angles θ and $\theta_D$ shown in Figs. 7c-d. $W_{eff}$ increases and $a/a_0$ decreases at the upstream side of the post when the RBC experiences increasing local shear rate while approaching the post top and then a recovery appears downstream, as shown in Figs. 7e-f.

At intermediate $Ca$, the fluid stresses are comparable to the elastic barrier for membrane tank-treading. We find $\theta_D$ to be locked around $-60^o$ before a discontinuous jump manifesting the effect of the elastic barrier occurs, as shown in Fig. 7c. A critical $Ca$ for going over the elastic barrier is between $Ca$=0.025 of the tumbling trilobe (TU-Tri in Fig. 7a) and $Ca$=0.1 of the tank-treading trilobe (TT-Tri in Fig. 7b). The recovery mechanism from high velocity gradients at the post top is schematically shown in Fig. 7a and Fig. 7b; the membrane material point moves backwards counter clockwise for TU-Tri, while for TT-Tri it goes across $\theta_D = 90^o$ and arrives at the dimple in the clockwise motion, following the flow.

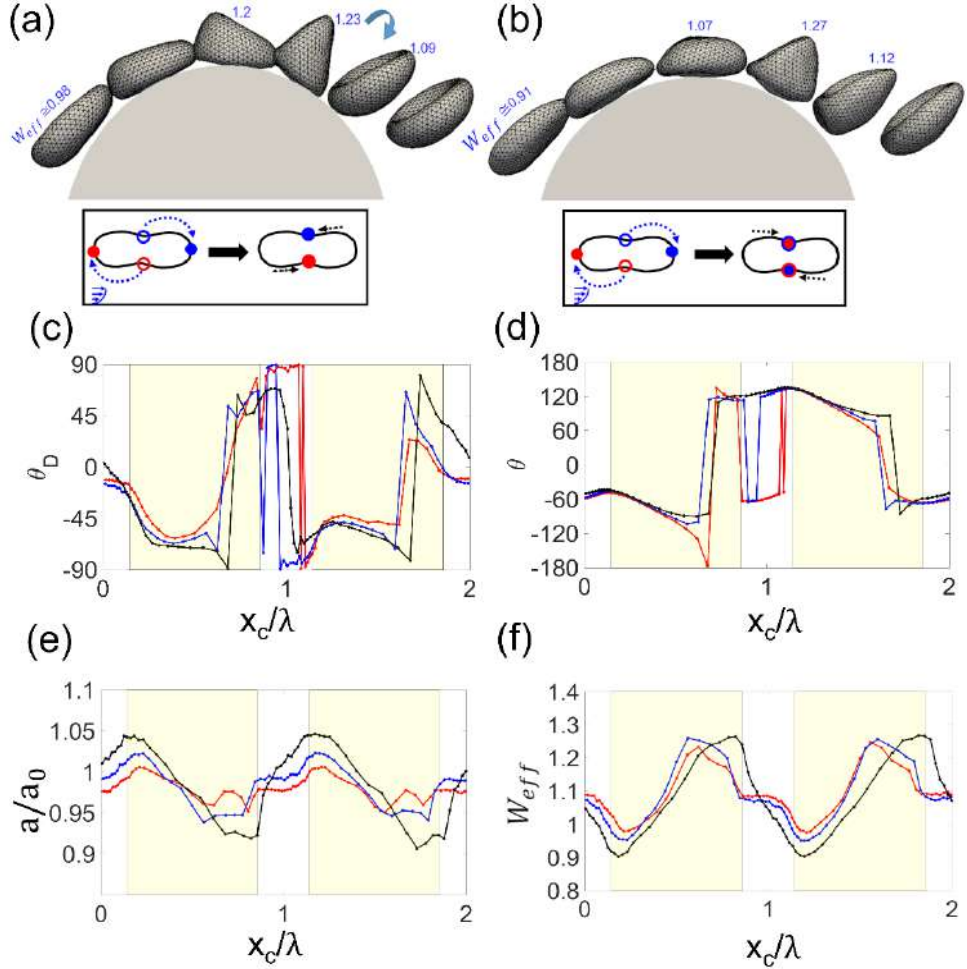

**FIG. 7.** RBC characteristics at the crossover between tumbling and tank-treading for intermediate Ca $\in [0.01; 0.15]$ in the DLD device with $\Delta\lambda = 2.0$ μm, $\Delta\lambda/\lambda = 0.057$. Snapshots for RBC dynamics at (a) Ca = 0.025 (Movie S2) and (b) Ca = 0.1 (Movie S3). The recovery processes of membrane material points are shown below, where open circles indicate initial locations and solid circles correspond to intermediate locations under flow. Evolution of (c) $\theta_D$, (d) θ, (e) $a/a_0$, and (f) $W_{eff} = bc/(b_0 c_0)$ for Ca = 0.025 (red), 0.05 (blue), and 0.1 (black). For a detailed description, see caption of Fig. 6.

**RBC behavior at high capillary numbers**

At high Ca > 0.15, the RBC membrane generally exhibits a TT motion, which is demonstrated by $\theta_D$ in Fig. 8b, spanning over 360 degrees. The inclination angle θ oscillates around an average value $\bar{\theta} \cong 112.5°$ (see Fig. 8c), which demonstrates that no TU motion occurs. Interestingly, the periodicity of $\theta_D$ (or membrane rotation) is not commensurate with the periodicity of DLD array structures. $a/a_0$ increases and $W_{eff}$ decreases as the RBC approaches the post top, where local shear rates are the highest, and these characteristics go in the reverse direction after passing the post top, as shown in Figs. 8d-e.

At Ca = 0.8, a marker located initially inside one of the dimples of the biconcave RBC shape (i.e. in the middle of the cell in the z-direction) moves to one of the sides of a tank-treading RBC, corresponding to a minimum or maximum in the z-direction perpendicular to the flow xy-plane (Fig. 9). Such a drift of membrane material points in flow occurs only at very high Ca, as shown in Fig. 9. The same membrane drift has also been observed for RBCs in simple shear flow.[45, 46] Another interesting observation is that the RBC does not directly touch post surfaces (Fig. 8a), as there is always a thin layer of fluid between the cell and the DLD posts (see Movie S4).

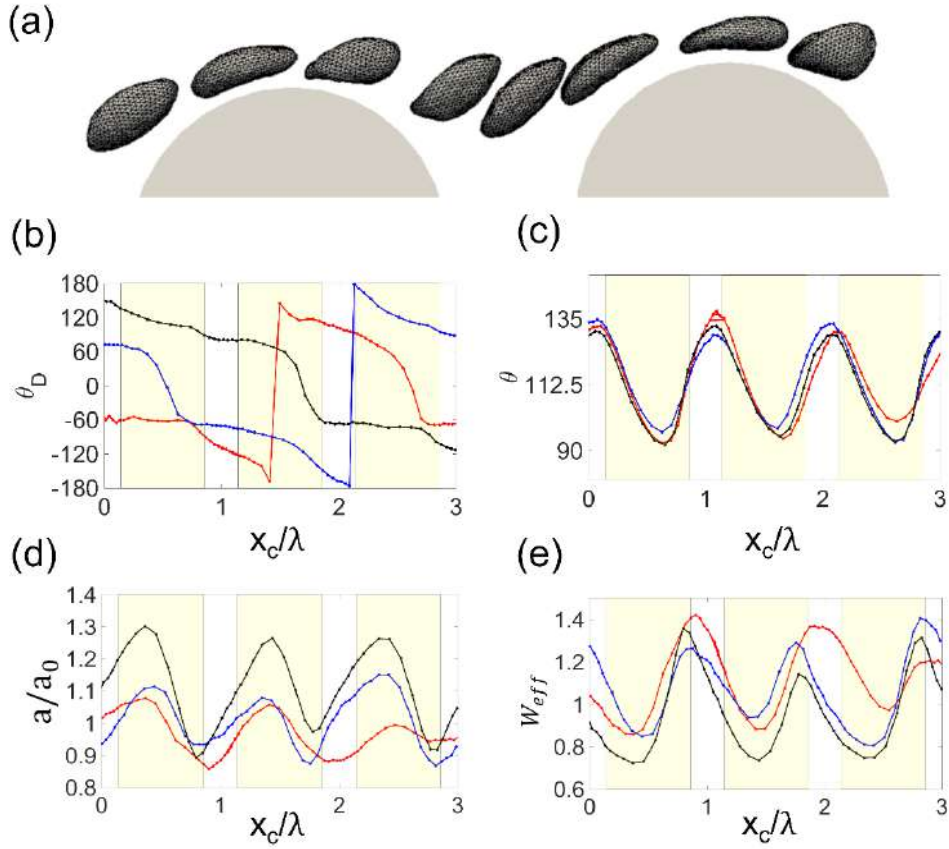

**FIG. 8.** Cell characteristics in a tank-treading dominated regime for high Ca > 0.15 in the DLD device with $\Delta\lambda$ = 2.0 μm, $\Delta\lambda/\lambda = 0.057$. (a) Snapshots for tank-treading dominated RBC dynamics at Ca = 0.4 (see also Movie S4). Evolution of (b) $\theta_D$, (c) θ, (d) $a/a_0$ and (e) $W_{eff}$ for Ca = 0.2 (red), 0.4 (blue), and 0.8 (black). For a detailed description, see caption of Fig. 6.

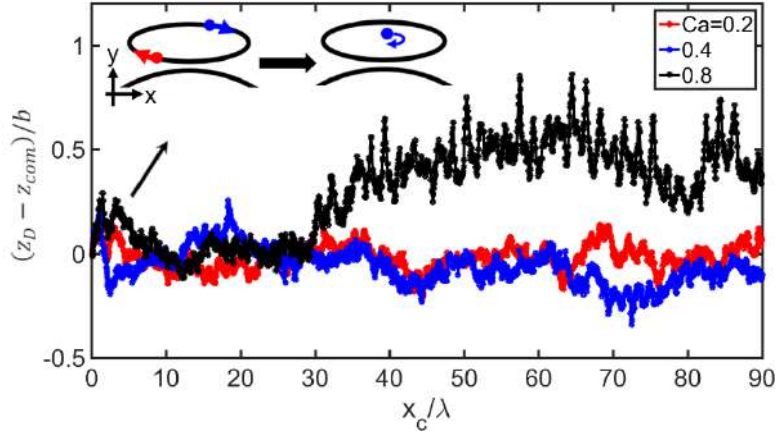

**FIG. 9.** The lateral drift of the membrane in z-direction for Ca = 0.8 (black) with relative membrane motion schematically shown. For Ca = 0.2 (red) and 0.4 (blue), no lateral drift occurs.

## C. The link between the cell's dynamic states and its effective size

To link RBC dynamics and deformation to the DP-ZZ transition in Fig. 2, we aim to determine an effective cell size for sorting from the complex dynamics. Figure 10 shows the deformation of the RBC as a function of Ca for DP modes. We find that the average shape of RBC remains the same below Ca = 0.01. For Ca = [0.01 0.15], we observe an increase of lateral effective stretching $W_{eff}$ and a compression of the long axis $a/a_0$ along the stream. For Ca > 0.2, the RBC elongates and $W_{eff}$ decreases. There is no direct correlation between the DP-ZZ transition boundary (or $\Delta\lambda_c$) and $W_{eff}$.

The key particle size to distinguish between DP and ZZ modes for deformable particles should be determined near the post top in relation to the width of the first stream. Following this idea, we calculate the maximum distance $d_{far}$ between the RBC outer membrane surface and the post surface near the post top, as shown in Fig. 11a. Indeed, $d_{far}$ is well correlated with $\Delta\lambda_c$. Furthermore, there is only the DP mode for $\Delta\lambda < 2.4\mu m$ and no DP mode for $\Delta\lambda > 3.6\ \mu m$ (see Fig.2), which implies that the effective size of the RBC is larger than $2\beta_{min}$ with $\beta_{min}$ being the width of the first stream for $\Delta\lambda = 2.4\mu m$ and smaller than $2\beta_{max}$, where $\beta_{max}$ is the width of the first stream for $\Delta\lambda = 3.6\mu m$. We obtain $2\beta_{min} = 3.03$ μm and $2\beta_{max} = 3.74$ μm from simulations without RBCs. Note that $d_{far}$ is exactly enclosed by $2\beta_{min}$ and $2\beta_{max}$ (see Fig. 11a), indicating that this range includes the effective size differences generated by RBC dynamics. Interestingly, $2\beta_{min}$ is larger than the thickness $2c_0 = 2.88\ \mu m$ of a RBC at rest.

At low Ca ≤ 0.01, the TU dynamics of the RBC effectively increases the critical thickness of the cell. This is illustrated well in Fig. 6a, where the RBC displays a flipping motion already at the top of the circular post. The clockwise flipping motion is against the first stream and along the outer streams, which favors the displacement mode. The flipping mechanism originates from a sterical collision between the RBC and the post which explains the fluctuating values of $I_d$ captured in Fig. 2.

At high Ca ≥ 0.2, $d_{far}$ increases as Ca increases (see Fig. 11a), even though $W_{eff}$ decreases (see Fig. 10b). Here, another physical mechanism, the hydrodynamic lift force, plays an important role,[47, 48] since a thin fluid layer is observed between the RBC and DLD posts in Fig. 8a. To quantify the thickness of the fluid layer separating the RBC and DLD posts, Fig. 11b displays the closest distance $d_{close}$ between the RBC membrane surface and the post surface. For Ca > 0.1, $d_{close}$ monotonically increases as a function of Ca, which is the main reason for an increase in $d_{far}$ in Fig. 11a.

In the intermediate regime, a minimal effective size appears. For Ca ∈ [0.01, 0.075), the RBC becomes more compressed as Ca increases. The compression reduces the extension away from the first stream and the destructive flipping effect mentioned above (see Fig.7a), and implies a reduction of the effective size. At higher Ca ∈ (0.075, 0.1], the contribution of partial TT motion becomes noticeable. This is due to passing of the elastic barrier which enhances tank-treading even when the RBC experiences decreasing local shear rates, as it moves away from the post top (see Fig. 7b). RBC tank-treading results in a significant hydrodynamic lift force, leading to an increase in the effective size as Ca increases.

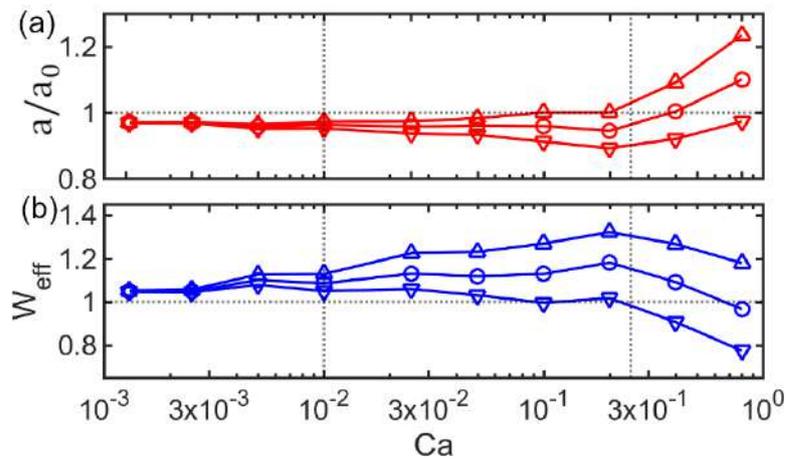

**FIG. 10.** Deformation of a RBC for DP modes with $\Delta\lambda$ = 2.0. Average (O), maximum (△), and minimum (▽) stretching $a/a_0$ (a) and effective lateral stretching $W_{eff}$ (b) of RBC from simulations with a cell displacement larger than $2n\lambda$ in the x-direction.

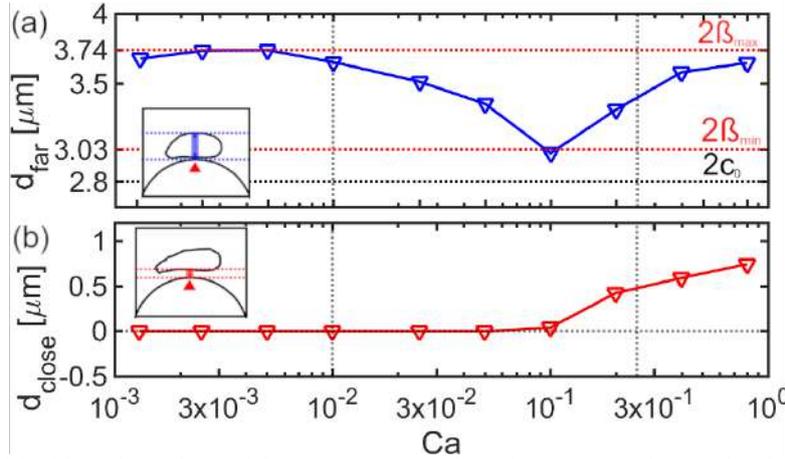

**FIG. 11.** Critical lengths estimated from DP modes with $\Delta\lambda = 2.0$. (a) The maximal distance $d_{far}$ between the RBC and the post surface near the post top, enclosed by the blue dashed lines in the inset. The red dotted lines mark the double thickness of the first stream $2\beta_{min} = 3.03$ μm at $\Delta\lambda = 2.4$ μm and $2\beta_{max} = 3.74$ μm at $\Delta\lambda = 3.6$ μm. The rest thickness $2c_0$ of the RBC is shown by the black dotted line. (b) The closest distance $d_{close}$ between the RBC and the post surface near the post tops, as shown in the inset. Both $d_{far}$ and $d_{close}$ are calculated over at least $2n$ post rows.

## IV. Summary and discussion

We have demonstrated a potential DLD device which can probe shear elasticity of RBCs by tuning $\Delta\lambda$ from 2.4 μm to 3.6 μm. The design of DLD devices with this range of $\Delta\lambda$ is realistic and has been realized in modern microfluidics.[16, 25] Compared to the device in Ref. 25 with $\lambda = 32$ μm, D = 20 μm, and G = 12 μm, our design with $\lambda = 35$ μm, D = 25 μm, and G = 10 μm has larger posts and smaller gaps, which provides a more parabolic velocity profile[7] with lower velocity gradients near post surfaces. This helps enhance the sorting contrast at different Ca, originating from the cell dynamics triggered near the post surface.

The detailed analysis of RBC deformation and dynamics demonstrates the complexity of cell behavior in a DLD device with circular post shape. Three different regimes have been identified. At low Ca $\leq 0.01$, the RBC exhibits TU dynamics with orientation change and small deformations. At intermediate $0.01 <$ Ca $\leq 0.15$, fluid stresses are large enough to induce partial TT motion of the membrane. As a result, the RBC shows a mixture of both TU and TT dynamics and experiences strong membrane deformations. At high Ca $> 0.15$, the cell attains TT motion. RBC dynamics in the DLD device is qualitatively similar to that in simple shear flow, where cell tumbling at low shear rates and tank-treading at high shear rates for low viscosity contrasts C are observed.[21, 28, 42, 43] The TU-TT transition is governed by the shape memory of a RBC,[28] which imposes an elastic energy barrier for TT motion.[42, 43]

The critical $\Delta\lambda_c$ at the DP-ZZ transition varies with Ca and cannot be simply described by the thickness of RBC shape in equilibrium. Instead, the dependence of $\Delta\lambda_c$ on Ca can be attributed to an effective RBC size, as the cell passes the top of posts. In fact, the effective cell size in the DLD device is larger than the thickness of the biconcave equilibrium shape for all Ca values. At low Ca, the RBC's effective size is governed by its flipping motion near the post top. At intermediate Ca, RBC deformations result in a decrease in the effective size, but this decrease stops when fluid stresses overcome the elastic barrier for membrane tank-treading. The elastic restoring force enhances tank-treading after this critical point and hydrodynamics lifting becomes important; thus, $\Delta\lambda_c$ begins to increase. At high Ca, the lift force increases, but the lateral cell size begins to decrease, resulting in a minor increase of $\Delta\lambda_c$. We expect that $\Delta\lambda_c$ at high Ca will reach a plateau since cell stretching nearly saturates at very high shear rates.

From the practical point of view, significant changes in $\Delta\lambda_c$ as a function of Ca are most interesting, as they allow the achievement of considerable contrasts in the lateral displacement of cells with distinct mechanical properties. For example, in the intermediate range of $0.01 < \text{Ca} \leq 0.1$, a choice of $\Delta\lambda \approx 2.6~\mu m$ should allow the separation of softer and stiffer RBCs representing different Ca values. Similarly, the range of $0.1 < \text{Ca} \leq 0.4$ is also characterized by a significant increase of $\Delta\lambda_c$ as a function of Ca, which can be exploited for sorting based on RBC elastic properties. However, it is important to emphasize that a careful selection of flow properties (e.g. flow rate, device geometry) is necessary in order to achieve efficient deformability-based sorting. Furthermore, the selected flow properties need to fit well the targeted elastic characteristics of RBCs, as both affect the capillary number.

Finally, it is important to discuss the limitations of our simulation study. The mode diagram in Fig. 2 represents RBC behavior in the DLD device at a viscosity contrast $C = 1$ between the suspending medium and RBC cytosol. Note that under physiological conditions the viscosity contrast is $C \approx 5$. In the regime of low Ca, we expect that a RBC with $C = 5$ would have a similar TU behavior as the RBC with $C = 1$. However, at high shear rates, deformation and dynamics of RBCs strongly depend on the viscosity contrast,[21, 44] such that the dependence of $\Delta\lambda_c$ at high enough Ca is likely to be different for $C = 5$ in comparison to that for $C = 1$. Nevertheless, Fig. 2 should represent well the behavior of RBCs with $C \leq 2$ in the investigated DLD device, because RBC dynamics in simple shear flow is independent of viscosity contrast in this regime.[21] Another limitation is that the variation in shear modulus has only been considered, while RBCs possess some variability in size and other mechanical properties such as bending rigidity. The variability in cell size is expected to have a moderate effect on the DP-ZZ transition, since RBCs expose their smallest size (i.e. thickness) due to cell orientation with $\Phi \approx 0°$ in the DLD device, and the variation in RBC thickness is not very large. Even though RBC shear elasticity is mainly responsible for the rich behavior of RBCs in simple shear flow,[21] the influence of membrane bending rigidity and of the stress-free shape on the DP-ZZ transition needs to be investigated in the future.

## V. Acknowledgments

We acknowledge the FP7-PEOPLE-2013-ITN LAPASO – "Label-free particle sorting" for financial support. The authors gratefully acknowledge the computing time granted through JARA-HPC on the supercomputer JURECA[49] at Forschungszentrum Jülich.